# Underwater Image Enhancement based on Deep Learning and Image Formation Model

Xuelei Chen[1], Pin Zhang[2], Lingwei Quan[1], Chao Yi[1], Cunyue Lu[1]

1. School of Electronic Information and Electrical Engineering, Shanghai Jiao Tong University, Shanghai 200240, China
2. National Key Laboratory on Electromagnetic Environmental Effects and Electro-optical Engineering, Nanjing, Jiangsu, 210007, China

**Abstract**：Underwater robots play an important role in oceanic geological exploration, resource exploitation, ecological research, and other fields. However, the visual perception of underwater robots is affected by various environmental factors. The main challenge now is that images captured by underwater robots are color-distorted. The hue of underwater images tends to be close to green and blue. In addition, the contrast is low and the details are fuzzy. In this paper, a new underwater image enhancement algorithm based on deep learning and image formation model is proposed. Experimental results show that the advantages of the proposed method are that it eliminates the influence of underwater environmental factors, enriches the color, enhances details, achieves higher scores in PSNR and SSIM metrics, and helps feature key-point point matching get better results. Another significant advantage is that its computation speed is much faster than other methods.

**Key words**：Deep learning; Image formation model; Image enhancement; Underwater robots; Visual perception

## 1 Introduction

Underwater robots rely on vision sensors to perceive their surroundings and make appropriate motion decisions when performing underwater tasks, and the underwater images they collect can also help and promote research in areas such as underwater biomes and geological changes. However, due to various environmental factors, the quality of underwater images is usually low. On the one hand, the overall color is green and blue; on the other hand, the contrast is low and the details are fuzzy. Therefore, it is of great importance to conduct research on underwater image enhancement methods.

There are many similarities between the physical image formation model of underwater images and fogged images [1], i.e., the light transmission is affected by the scattering of particles in the air or water, which causes the light entering the camera to be different from the light reflected from the object itself, thus degrading the captured images. Researchers have proposed various methods [2-3] for underwater image color correction by directly applying the physical image formation model of fogged images to underwater images. However, the propagation of light in the water is different from that in the atmosphere. The existing image formation models do not consider the difference between the scattering effect of light with different wavelengths. In the latest paper by Akkaynak et al. [4], the physical image formation model of underwater images has been further studied. Through a large number of real underwater experiments, the attenuation coefficient of the existing model was modified, and a revised underwater image formation model was proposed.

In recent years, deep learning has achieved very good results on computer vision tasks such as image recognition, image segmentation, and object detection. More and more researchers begin to apply deep learning to underwater image enhancement tasks. Li et al. [5] proposed an underwater image color correction method based on CycleGAN, Skinner et al. [6] proposed a two-stage neural network structure for image depth estimation and color correction respectively, and Fu et al. [7] proposed a neural network combining global and local information for underwater image enhancement.

The above methods based on physical imaging models are often only applicable to specific images and are not generalizable, and most of the existing methods are based on approximate image formation models. Some of the deep learning-based methods ignore the image formation model, which leads to a complex model structure and makes training process more difficult. Some other deep learning-based methods incorporate the image formation model but simplify it, which can lead to insignificant image enhancement effects.

In this paper, a new method for underwater image enhancement is proposed by improving the existing deep learning-based image enhancement method by combining the recently proposed revised image formation model [4]. The proposed method consists of a backscatter estimation module and a direct-transmission estimation module, both of which are implemented using convolutional neural networks. The outputs of these two modules together with the input image are fed into a reconstruction module to obtain the enhanced underwater image. The parametric rectified linear unit (PReLU) [8] and the dilated convolution [9] are used to improve the fitting ability when we construct the neural network. Experiments on UIEB [10] dataset and URPC2020 [11] dataset validate the effectiveness of the proposed method. The code and the trained model of the proposed method are open-sourced on GitHub:
https://github.com/xueleichen/PyTorch-Underwater-Image-Enhancement

## 2  Physical Image Formation Model
### 2.1 Optical Imaging Principles

When taking photographs in underwater environments, the image quality is affected by factors such as light absorption, light refraction, and light scattering [12]. These factors can cause different effects on underwater images.

Light will be absorbed when passing through the water, and the absorption degree of light with different wavelengths is different. Figure 1 shows the process of light absorption in underwater environments. The seawater has a stronger absorption effect on the red light, which rapidly decreases in intensity after entering the body of water，while green and blue light with shorter wavelengths can penetrate deeper below the sea level, resulting in a greenish and blueish effect in images taken underwater.

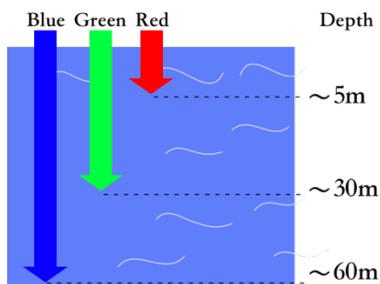

**Fig.1 Illustration of underwater light absorption**

The optical image formation model of underwater images consists of two parts of light source information: the directly transmitted light and the background scattered light. As shown in Figure 2, the directly transmitted light comes from the object itself, and it will experience an attenuation effect on the transmission path. The attenuation effect is caused by light absorption and light scattering, which will cause color distortion of underwater images. The background scattered light does not originate from the radiation of the object itself but is caused by the light in the surrounding environment being scattered by a large number of tiny particles in the water. the background scattered light is the main reason for the decrease in the contrast of underwater images.

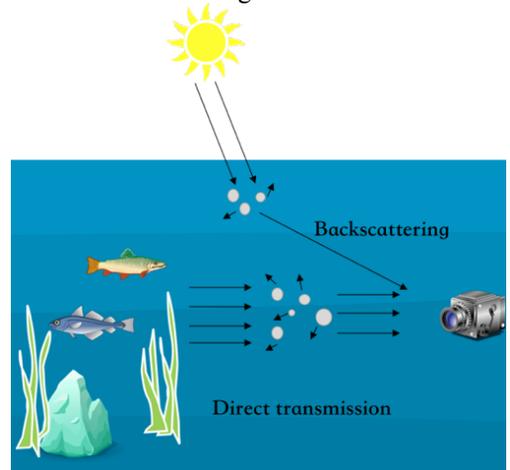

**Fig.2 Illustration of optical imaging principles**

In addition, light does not travel in a straight line in an inhomogeneous medium and is subject to refraction. Refraction of light can cause distortion of details in the image [13]. Particles, bubbles, and the fluctuation of seawater can cause refraction of light thus causing distortion in underwater images. Since the recovery of this kind of image distortion belongs to the field of image restoration, the method proposed in this paper belongs to the image enhancement, which only solves the problems of color distortion, low contrast, and blurred details caused by light absorption and light scattering.

### 2.2 Mathematical Formulation

An approximate underwater image formation model [2] can be obtained from the image formation model of fogged images [1]：

$$I(x) = D(x)t(x) + B(1 - t(x)) \quad (1)$$

where $x$ denotes the coordinates of the pixel, $I(x)$ denotes the image obtained by the camera in the underwater environment, $D(x)$ denotes the radiation of the object itself，which can be understood as the underwater image after eliminating the influence of underwater environmental factors, $t(x)$ is the direct-transmission mapping，$B$ is the ambient light. $D(x)t(x)$ corresponds to the directly transmitted light，and

$B(1-t(x))$ corresponds to the background scattered light, as shown in Figure 2.

The directly transmitted light will experience an attenuation effect，the magnitude of which is determined by the attenuation coefficient $\beta$ and the transmission distance $d$, as formulated in the following equation：

$$t(x) = e^{-\beta d} \quad (2)$$

The existing methods for underwater image enhancement usually treat the $t(x)$ of three channels of an RGB image as one mapping to simplify the calculation，but this approach ignores the difference of the attenuation coefficient $\beta$ in $t(x)$ between different channels and the enhanced image still has color bias. Akkaynak et al. [4] conducted a large number of underwater experiments and proposed a revised model, as shown in Equation (3). The revised model modifies the attenuation coefficient based on the optical imaging characteristics of the underwater environment.

$$I_c(x) = D_c(x)e^{-\beta_c^D(V_D)d} + B_c\left(1 - e^{-\beta_c^B(V_B)d}\right) \quad (3)$$

where $c \in \{R, G, B\}$ denotes the channel, $\beta_c^D$ 和 $\beta_c^B$ denote the attenuation coefficients for direct-transmission and backscatter, and $d$ denotes the transmission distance. The attenuation coefficients of direct-transmission and backscatter depend on $V_c^D = [d, \rho, E, S_c, a, b]$ and $V_c^B = [E, S_c, a, b]$, where $d$ is the transmission distance, $\rho$ is the reflection spectrum, $E$ is the irradiance, $S_c$ is the spectral response of the camera, $a$ and $b$ are the absorption and scattering coefficients, respectively.

Organizing Equation (3), we can obtain the following Equation (4):

$$D_c(x) = (I_c(x) - B_c)e^{\beta_c^D d} + B_c e^{(\beta_c^D - \beta_c^B)d} \quad (4)$$

According to the experiments in [5], the value of $\beta_c^D - \beta_c^B$ is very small and can be approximated as 0 at distances greater than 3m, so Equation (4) can again be approximated as the form of Equation (5):

$$D_c(x) = (I_c(x) - B_c)e^{\beta_c^D d} + B_c \quad (5)$$

This paper use Equation (5) as the physical image formation model of underwater images, and combine it with deep learning techniques to complete the underwater image enhancement task.

## 3 Deep Learning Method

The core idea of the proposed method is to use convolutional neural networks to fit multiple components in the image formation model. As shown in Figure 3, the purposed method consists of three parts：the first part estimates the ambient light from the input image, that is, $B_c$ in Equation (5). The second part uses the estimated ambient light and the input image to estimate the direct transmission map, that is $e^{\beta_c^D d}$ in Equation (5). Here, unlike most existing methods to estimate $t(x)$, $e^{\beta_c^D d}$ is actually the inverse of $t(x)$. The advantage of this setting is that the final reconstruction operation is multiplicative, which facilitates the backpropagation of the loss function in the deep learning training. The third part uses the estimated $B_c$ and $e^{\beta_c^D d}$ combined with Equation (5) to perform the reconstruction operation to obtain the enhanced image, in which the influence of the underwater environment is eliminated.

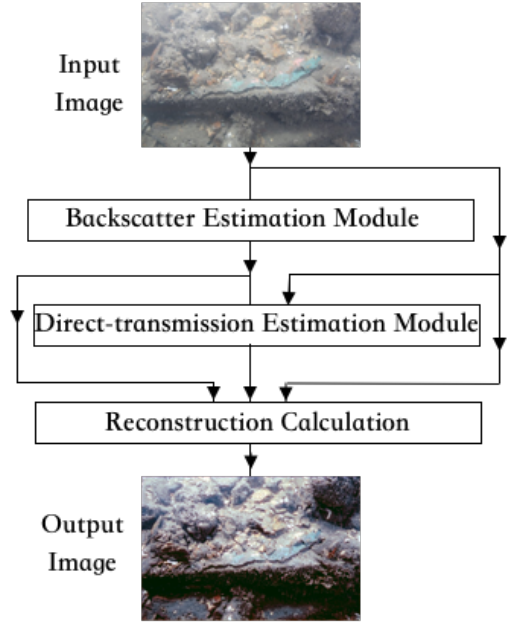

**Fig.3 Framework of the proposed method**

### 2.1 Backscatter Estimation Module

The backscatter estimation module includes two groups of 3x3 convolution kernels, one global mean pooling layer and two groups of 1x1 convolution kernels. The number of each group of convolution kernels is 3. The detailed structure is shown in Figure 4. The parametric rectified linear unit (PReLU) [9] is chosen as the activation function of the convolution operation.

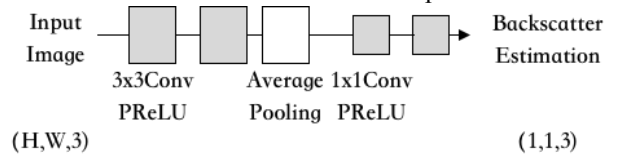

**Fig.4 Backscatter estimation module**

The parametric rectified linear unit (PReLU) and the rectified linear unit (ReLU) is formulated in Equation (7) and (8)：

$$\text{PReLU}(x) = \begin{cases} x, & \text{if } x \geq 0 \\ ax, & \text{otherwise} \end{cases} \quad (7)$$

$$\text{ReLU}(x) = \begin{cases} x, & \text{if } x \geq 0 \\ 0, & \text{otherwise} \end{cases} \quad (8)$$

where $x$ is the value of each entry on the feature map in the neural network and $a$ is a learnable parameter. Compared with the normal rectified linear unit (ReLU)，the parametric rectified linear unit (PReLU) avoid the phenomenon that the gradient becomes zero when $x$ is negative and the weights can no longer be updated。

## 2.2 Direct-transmission Estimation Module

The direct transmission estimation module first concatenates the backscatter estimation with the input image, and the subsequent operation includes three groups of 3x3 dilated convolutional kernels and one group of 3x3 normal convolutional kernels. The number of each group of dilated convolution kernels is 8, and the number of the last group of normal convolution kernels is 3. The detailed structure is shown in Figure 5.

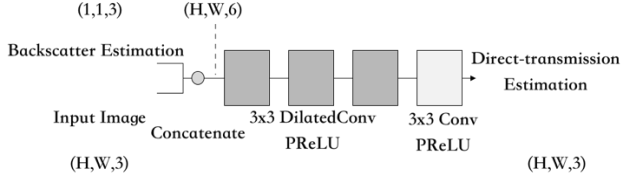

**Fig.5 Direct-transmission estimation module**

The dilated convolution can enlarge the receptive field by filling extra zeros in the convolution kernel without adding more parameters. In order to avoid the gridding effect [10], we can adopt the design method of hybrid dilated convolution, that is, different dilation rates are selected for consecutive dilated convolution operations. The dilation rates of three groups of dilated convolution kernels in the module are respectively [1,2,5].

## 2.3 Training method

The output of the proposed method is the enhanced image. In the training process, each input image has a corresponding reference image as the training target. The mean square error based on the pixel value is used as the loss function to calculate the difference between the enhanced image and the target image. The mean square error loss function is shown in Equation (6):

$$L = \frac{1}{HWC} \sum_{x,y,z=1}^{H,W,C} \left(\widehat{D}(I)_{x,y,z} - R(I)_{x,y,z}\right)^2 \quad (6)$$

where $H$, $W$ and $C$ are the height, width and number of channels of the image, respectively, $x$, $y$ and $z$ denote the coordinates of the three dimensions of the image, $\widehat{D}(I)$ denotes the output of the proposed method, and $R(I)$ denotes the reference image in the training dataset。Adam optimizer [14] is used to optimize neural network parameters.

## 3 Experiments

In order to verify the effectiveness of the underwater image enhancement method proposed in this paper, a series of experiments are designed and conducted. The experiments were conducted in Ubuntu 18.04 system using PyTorch deep learning framework. The computer used has Intel i7-10700 CPU and NVIDIA RTX 2070 GPU.

The batch size used in the deep learning training is 1, the learning rate is 0.001, and the number of epochs is 3000. The source of the images in the deep learning experiments is the UIEB [11] dataset, which collects a large number of underwater images taken in real scenes. This dataset also contains corresponding color-enhanced images obtained by a variety of existing algorithms and manual evaluation of volunteers. The dataset contains 890 real underwater images and 890 color-enhanced images for reference, of which 700 images are used for training and 190 images are used for testing. In addition, the model trained on the UIEB dataset was also tested on the URPC2020 dataset to verify the generality of the proposed method. In the analysis of experimental results, the effectiveness is evaluated in terms of visual quality, quantitative metrics and computation speed. And the results of the proposed method are compared with state-of-the-arts. What's more, experiments on feature point matching are conducted to show that underwater image enhancement can boost high-level computer vision tasks.

### 3.1 Qualitative Evaluation

For visual quality, experimental results of the proposed method are compared with the results of several state-of-the-art methods including UDCP [2], IBLA [3], and GLNet [8], as shown in Figure 6. The images in the first three rows of Fig. 6 are from the test set of UIEB dataset, and the images in the last three rows are from the URPC2020 dataset. the first column is the original image of the input, and the last four columns are the results obtained by using different methods. From Figure 6, we can see that the proposed method can achieve better results when performing underwater image enhancement. It can correct the color bias of underwater images, improve the contrast and detail clarity, and enhance the overall visual effect. Other methods will have dark background or overall bluish after correction.

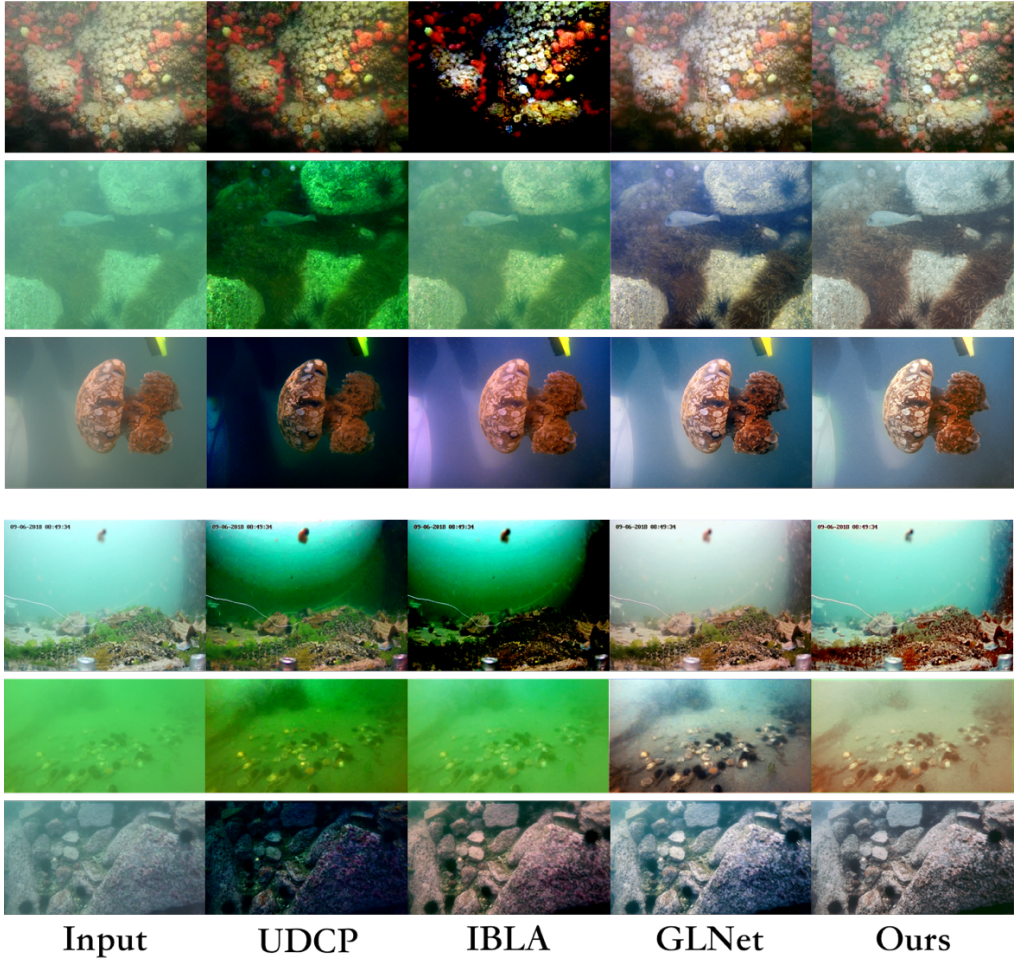

**Fig. 6 Visual comparison of different underwater image enhancement methods**

## 3.2 Quantitative Evaluation

Quantitative metrics used in underwater image enhancement experiments can be divided into two categories: full-reference metrics and non-reference metrics. Calculating full-reference metrics requires that the image processed by the algorithm has a corresponding reference image, thus enabling the calculation of the approximation between these two images. Commonly used full-reference metrics include peak signal-to-noise ratio (PSNR) and structural similarity index (SSIM). Non-reference metrics directly use the single processed image to calculate the color, contrast, saturation, etc. and then combine them into one overall metric. Commonly used non-reference metrics include underwater color image quality evaluation (UCIQE) [15] and underwater image quality evaluation (UIQM) [16].

Table 1 shows the values of each metric on the UIEB test set for the proposed method and UDCP [2], IBLA [3], and GLNet [8]. Compared with GLNet, which is also based on deep learning, the proposed method performs better in the PSNR and SSIM metrics and slightly worse in the UCIQE and UIQM metrics. The reason is that GLNet method uses histogram equalization as the post-processing step, which makes the image color more balanced and achieves higher values in the metrics. However, it also leads to the problem that some images processed by GLNet are bluish, as shown in Figure 6. The visual quality of images processed by traditional algorithms such as UDCP and IBLA is not good, but they get the highest score in UCIQE. This is because UCIQE pays more attention to contrast. Even if the image is excessively enhanced, resulting in color deviation, it is still possible to get a higher UCIQE value. This phenomenon is also discussed in paper [11], which shows that the current underwater image quality metrics still have limitations.

**Table 1 Comparison of different underwater image enhancement methods by quantitative metrics**

|      | PSNR  | SSIM  | UCIQE | UIQM  |
|------|-------|-------|-------|-------|
| UDCP | 20.23 | 0.470 | 9.539 | 2.414 |
| IBLA | 19.91 | 0.654 | 7.296 | 3.283 |
| GLNet | 19.36 | 0.873 | 5.468 | 4.196 |
| Ours | 19.37 | 0.877 | 4.899 | 3.975 |

### 3.4 Computation Speed

When the underwater image enhancement method is deployed to real robots, the real-time performance of the algorithm should also be considered. Here, the computation speed of the proposed method is compared with UDCP [2], IBLA [3] and GLNet [8]. The specific results are shown in Table 2. The comparison here is in terms of fps (frames per second). We can see that the proposed method has the fastest computation speed. The reason for the high speed is that the number of convolutional kernels per group is chosen to be small when designing the neural network, which makes the network lighter.

**Table 2 Comparison of different underwater image enhancement methods by computation speed**

|  | UDCP | IBLA | GLNet | Ours |
|---|---|---|---|---|
| fps | 0.041 | 0.018 | 0.471 | 9.868 |

### 3.5 Boosting High-level Vision Tasks

The purpose of underwater image enhancement is to improve the visual perception ability of underwater robots to the surrounding environment. Image feature point matching is a basic visual perception task, which is the basis of photogrammetry, 3D reconstruction and image stitching.

Firstly, the proposed method is used to enhance two images numbered 000001 and 000002 in URPC2020 dataset, and then feature points of two original images and two enhanced images are matched respectively. SIFT method [17] is used to extract feature points, RANSAC method [18] is used to match feature points and homography transformation. The experimental results are shown in Figure 7, the figure on the left is 000001, and the figure on the right is 000002. It can be seen that more matching points can be detected in the enhanced image, which also proves the effectiveness of the image enhancement.

Meanwhile, the above two images were also processed using different underwater image enhancement methods and then matched using SIFT and RANSAC. The numbers of matched points are compared in Table 3. It can be observed that the proposed method can boost feature point matching the most.

**Table 3 Comparison of feature point matching on images enhanced by different algorithms**

|  | UDCP | IBLA | GLNet | Ours |
|---|---|---|---|---|
| # pairs | 3726 | 5689 | 7411 | 8633 |

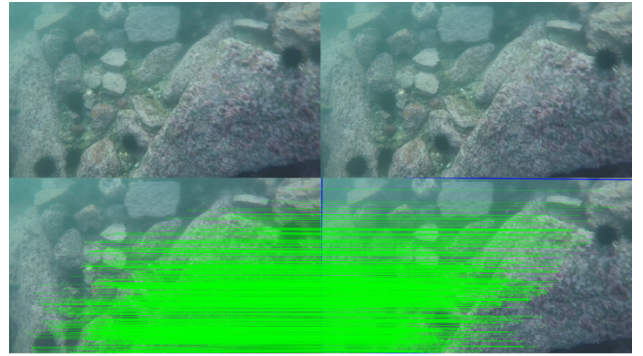

(a) Original images: 1384 pairs of matched points

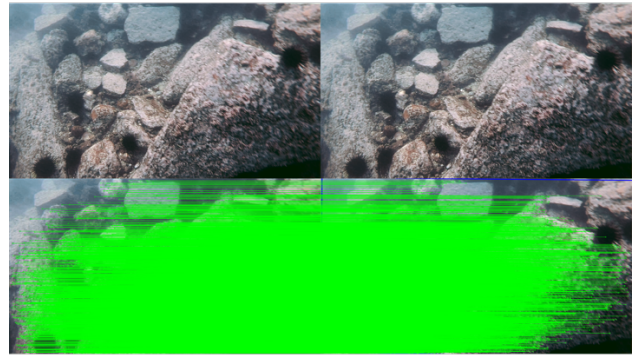

(b) Enhanced images: 8633 pairs of matched points

**Fig.7 Feature key-point matching experiments**

## 4 Conclusion

In this paper, a new underwater image enhancement method based on deep learning and image formation model is proposed. The parametric rectified linear unit (PReLU) [8] and the dilated convolution [9] are used to improve the fitting ability when we construct the neural network. The proposed method achieves good image enhancement results on several datasets, with improvements in both visual quality and quantitative metrics compared to existing methods. At the same time, the computation speed is greatly improved, which can meet the real-time computing requirements of the real underwater robot platform in the future. Moreover, experiments show that the proposed underwater image enhancement methods can assist subsequent high-level computer vision tasks.

We will continue to explore the application of image enhancement in other computer vision tasks such as underwater object detection, and try to replace the current supervised learning with unsupervised learning. In addition, this paper only uses the information of the image itself, and in the future, we will integrate the spatial perception information of various sensors such as echo sounder and Doppler rangefinder for more accurate underwater image color correction and image enhancement. For distortion caused by the refraction of light, we will further improve the current method to realize the distortion restoration of underwater images along with image enhancement in one framework.

**Reference:**

[1] Narasimhan S G, Nayar S K. Vision and the Atmosphere[J]. International Journal of Computer Vision, 2002, 48(3):233-254.
[2] Drews Jr. P, do Nascimento E, Moraes F, et al. Transmission Estimation in Underwater Single Images[C]. Proceedings of the International Conference on Computer Vision Workshops, 2013:825-830.
[3] Peng Y, Cosman P C. Underwater Image Restoration Based on Image Blurriness and Light Absorption[J]. IEEE Transactions on Image Processing, 2017, 26(4):1579-1594.
[4] Akkaynak D, Treibitz T. A Revised Underwater Image Formation Model[C]. Proceedings of the IEEE Conference on Computer Vision and Pattern Recognition, 2018:6723-6732.
[5] Li C, Guo J, Guo C. Emerging From Water: Underwater Image Color Correction Based on Weakly Supervised Color Transfer[J]. IEEE Signal Processing Letters, 2018 25(3):323-327.
[6] Skinner K A, Zhang J, Olson E A, et al. UWStereoNet: Unsupervised Learning for Depth Estimation and Color Correction of Underwater Stereo Imagery[C]. Proceedings of the International Conference on Robotics and Automation, 2019:7947-7954.
[7] Fu X, Cao X. Underwater image enhancement with global–local networks and compressed-histogram equalization[J]. Signal Processing: Image Communication, 2020, 86: 115892.
[8] He K, Zhang X, Ren S, et al. Delving deep into rectifiers: Surpassing human-level performance on imagenet classification[C]. Proceedings of the International Conference on Computer Vision, 2015:1026-1034.
[9] Wang P, Chen P, Yuan Y, et al. Understanding convolution for semantic segmentation[C]. Proceedings of IEEE Winter Conference on Computer Vision, 2018:1451-1460.
[10] Li C, Guo C, Ren W, et al. An Underwater Image Enhancement Benchmark Dataset and Beyond[J]. IEEE Transactions on Image Processing, 2019, 29:4376-4389.
[11] Underwater Object Detection Algorithm Contest[OL]. http://uodac.pcl.ac.cn/
[12] Hitam M S, Awalludin E A, Yussof W N J H W, et al. Mixture Contrast Limited Adaptive Histogram Equalization For Underwater Image Enhancement[C]. Proceedings of the International Conference on Computer Applications Technology, 2013:1-5.
[13] Tian Y, Narasimhan SG. Seeing throu water: Image restoration using moder-based tracking[C]. Proceedings of the International Conference on Computer Vision, 2009: 2303-2310.
[14] Kingma D P, Ba J. Adam: A Method for Stochastic Optimization[C]. Proceedings of the 3rd International Conference for Learning Representations, 2015.
[15] Yang M, Sowmya A. An underwater color image quality evaluation metric[J]. IEEE Transactions on Image Processing, 2015,24(12):6062-6071.
[16] Panetta K, Gao C, Agaian S. Human-visual-system-inspired underwater image quality measures[J]. IEEE Journal of Oceanic Engineering, 2015,41(3):541-551.
[17] Lowe D G. Distinctive image features from scale-invariant keypoints[J]. International Journal of Computer Vision, 2004,60(2):91-110.
[18] Fischler M A, Bolles R C. Random sample consensus: a paradigm for model fitting with applications to image analysis and automated cartography[J]. Communications of the ACM, 1981,24(6):381-395.